# OpenSPIM - an open access platform for light sheet microscopy

Peter Pitrone[1,*], Johannes Schindelin[2,*], Luke Stuyvenberg[2], Stephan Preibisch[1,3], Michael Weber[1], Kevin W. Eliceiri[2], Jan Huisken[1], Pavel Tomancak[1]

[1]Max Planck Institute of Molecular Cell Biology and Genetics, Pfotenhauerstraße 108, Dresden, Germany

[2]Laboratory for Optical and Computational Instrumentation, University of Wisconsin at Madison, 1675 Observatory Drive, Madison, WI 53706, USA

[3]Albert Einstein College of Medicine, 1300 Morris Park Avenue, Bronx, NY 10461, USA

*these authors contributed equally

Correspondence should be addressed to: tomancak@mpi-cbg.de

**Abstract**

**Light sheet microscopy promises to revolutionize developmental biology by enabling live *in toto* imaging of entire embryos with minimal phototoxicity. We present detailed instructions for building a compact and customizable Selective Plane Illumination Microscopy (SPIM) system. The integrated OpenSPIM hardware and software platform is shared with the scientific community through a public website, thereby making light sheet microscopy accessible for widespread use and optimization to various applications.**

Light sheet microscopy, such as SPIM[1], overcomes many of the current limitations of common confocal and multiphoton microscopes: high phototoxicity, low acquisition speeds and poor depth penetration. In SPIM only a single slice of the sample is illuminated at a given time reducing phototoxicity to a minimum. This slice is imaged onto a fast and highly sensitive camera and a three-dimensional volume of several hundreds of micrometers is imaged in a few seconds. To resolve all parts of a millimeter sized sample equally well the SPIM features a unique rotational motor to turn the sample and image it from multiple directions. These features of SPIM make the instrument highly desirable for the growing group of life scientists studying phenomena in entire tissues, organs and embryos.[2-4]

Recent developments in the optical construction of light sheet microscopes make it possible to capture every cell in the embryo throughout development resulting in large scale reference recordings of embryo anatomy.[5-8] Reverse genetic techniques in major model organisms have matured to the point that fluorescent gene expression reporters can be generated on a genome-wide scale.[9-11] SPIM is ideally suited to capture definitive patterns of gene expression in the context of full embryo anatomy using these reporters. However, since animal genomes contain tens of thousands of genes, development can last many hours or days and since experiment replication is paramount for proper interpretation, it is necessary to increase the throughput of SPIM acquisitions. One option is to deliver multiple samples into the microscope's field of view repeatedly, however such approaches necessarily compromise the temporal resolution of live imaging. We propose here to increase the acquisition throughput by building arrays of affordable special-purpose SPIM systems allowing parallel imaging of many samples. Additionally, by employing open principles in both hardware and software (e.g. Wiki style documentation, open designs for 3d-printable parts, source code versioning, etc.) and by providing detailed instructions that do not require prior expertise in building optical setups we make the OpenSPIM platform accessible to any laboratory for biological applications and further technical development.

The initial version of the OpenSPIM setup implements a single sided illumination and single sided detection light sheet setup (**Fig. 1a-i,a'-i'**), the most basic realization of the SPIM principle, ready to be optimized for particular imaging tasks. The entire setup fits onto an optical breadboard of 30 x 45 cm (1' x 1'6") (**Fig. 1a**). In its simplest form the OpenSPIM offers one 488 nm laser line (Coherent Cube, Santa Clara, USA) (**Fig. 1b**). The illumination axis creates the light sheet as described previously (**Fig. 1c-f**).[1,12] The detection axis consists of a 20x/0.5 water dipping objective (Olympus, Tokyo, Japan) attached via a custom made cylindrical spacer to a tube lens, c-mount, and a CCD camera (Hamamatsu Orca ER, Japan) (**Fig. 1g-i**). Emission filters can be easily inserted into a slit in the spacer (**Fig. 1j**). The illumination and detection arms meet at 90 degrees angle in the water-filled sample chamber machined from acrylic according to the provided plans (**Fig. 1k**). The 4D positioning system (Picard industries, Albion, USA) controls rotation and translation of the sample through the light sheet within the chamber (**Fig. 1l**). The sample is mounted in agarose and extruded from a capillary fixed in position by a cut plastic syringe serving as sample holder (**Fig. 1m**). The entire setup can be mounted inside a rack, or even inside a cabin sized suitcase.

The OpenSPIM is controlled via the popular Open Source microscopy control software µManager running on standard computer hardware (**Fig. 1n-p**).[13] We extended µManager to support the 4D positioning stage and to enable enhanced computer controlled sample positioning (**Fig. 1o**) (in addition to µManager's three-dimensional translation support via dragging or scrolling the mouse inside the live view, OpenSPIM also supports interactive rotation by moving the mouse while holding down the Alt key), synchronized laser and camera triggering and multi-view time-lapse acquisition (**Fig. 1p**). OpenSPIM's software operates as a µManager plugin and uses Fiji's software components for hardware control triggered by advanced image analysis (**Fig. 1n**). The data for each acquired view are stored in separate OME-Tiff[14] files available for processing in any Open Microscopy Environment (OME) compliant software. Together with µManager, the OpenSPIM acquisition software is embedded in ImageJ's distribution Fiji[15] where the acquired data are available for bead based multi-view reconstruction and fusion[16] as well as further analysis (**Fig. 1n**). Additionally, we use Fiji's update mechanism to distribute and update the OpenSPIM software components via OpenSPIM's own update site (**Fig. 1n**). The combination of OpenSPIM open access hardware and open source software form an accessible and highly adaptable platform for multi-view light sheet imaging in biology (**Fig. 1**).

To demonstrate the ability of OpenSPIM to image large specimen we imaged a two days old zebrafish embryo expressing H2A-eGFP in all cells by acquiring six separate fields of view from one angle and stitching them together in Fiji (**Fig. 2a** and **Supplementary Video 1**).[17] OpenSPIM can capture fast biological processes such as beating zebrafish heart by acquiring a single plane in a continuous movie mode at the maximum frame rate of the camera (**Fig. 2b** and **Supplementary Video 2**). For strongly scattering samples, such as *Drosophila* embryos, the OpenSPIM offers a multi-view acquisition mode. We imaged *Drosophila* embryos expressing Histone YFP (His-YFP) in all cells from 6 views and reconstructed the data using bead based SPIM registration algorithm[16] in Fiji resulting in complete coverage of the specimen. The contrast of the data can be dramatically improved by multi-view deconvolution also available in Fiji (Preibisch et al. manuscript in preparation) (**Fig. 2c,d** and **Supplementary Video 3 and 4**). A notorious problem of many SPIM time-lapse acquisitions is the drift of the sample out of the microscope's field of view over time due to agarose or mechanical instabilities of the system. To counteract that problem we have developed an algorithm that detects the sample drift robustly and keeps the specimen in the field of view by adjusting the motor position during acquisition (http://openspim.org/Anti-Drift). Using the sample drift correction we were able to record a long-term multi-view time-lapse of *Drosophila* embryos expressing His-YFP from gastrulation until muscle contraction (**Fig. 2e** and **Supplementary Video 5**) and Csp-sGFP protein fusion expressed from a third copy tagged allele[10] (**Fig. 2f** and **Supplementary Video 6**). The 4D recording of the dynamics of gene expression pattern formation during development represents the major application where parallelization offered by OpenSPIM provides the necessary throughput.

The unique feature of the OpenSPIM project is its emphasis on an open development process and community. We have established a wiki site (http://openspim.org) which documents in detail the building, alignment, software and operation of the OpenSPIM microscope. The OpenSPIM wiki contains a comprehensive list of commercially available parts necessary to build the setup, as well as CAD drawings of a few custom parts that cannot be purchased but need to be made by local mechanical workshops or 3D printed (http://openspim.org/Table_of_parts). Detailed instructions for step-by-step assembly enriched with photographs, CAD drawings and numerous video tutorials allow researchers even without prior experience in optical technology to build the OpenSPIM setup (http://openspim.org/Step_by_step_assembly). Narrated videos guide the users through the initial steps of light sheet alignment and sample preparation[18,19] (http://openspim.org/Operation). The Open Source software that drives OpenSPIM is documented with tutorials for the casual users as well as in depth explanations of installation and the software engineering principles for more experienced developers interested in further modifying and extending the platform (http://openspim.org/Category:Software). In the wiki environment anybody can register, edit and add pages and thus contribute to further development of the OpenSPIM concept.

The OpenSPIM was intentionally designed to be simple, compact, modular and accessible. The current OpenSPIM is limited to one sided illumination and detection (**Fig. 3a**), however we intend to extend the original design to more advanced SPIM geometries (two sided illumination and/or detection) (**Fig. 3b,c**) as well as more sophisticated light sheet formation paradigms (pivoting[12], DSLM[20]), multi-color imaging and more advanced cameras and positioning systems. The modular properties of OpenSPIM will make it possible to multiplex long-term time-lapse acquisitions by building several OpenSPIM setups operating in parallel (SPIM farm) (**Fig. 3d**). Its simplicity makes it an excellent teaching tool, both for applying and building SPIM technology; OpenSPIM's components can be easily transported to a course or conference and built on site with the students using the wiki instructions. The combination of open access hardware and Open Source software make OpenSPIM an ideal platform to adapt for specific imaging tasks as well as for rapid prototyping of new SPIM configurations. Most importantly the OpenSPIM and its associated wiki is an open platform that invites interested researchers to build the setup, to improve upon it on various levels and to contribute the experience and innovations to the scientific community. We envision that this will nucleate an interdisciplinary community where biologists interested in using SPIM imaging will interact with technology developers to bring to life innovative ideas for addressing complex scientific questions.

**Acknowledgements**

We would like to thank Vineeth Surendranath for help with photography, Hugo Bellen for the Csp-sGFP transgene, Sukhdeep Singh, Sonal and Steve Simmert for seeding the wiki with material during the MPI-CBG PhD course. P.T. and P.P. were supported by The European Research Council Community's Seventh Framework Programme (FP7/2007-2013) grant agreement 260746. S.P. was supported by the Human Frontier Science Program (HFSP) Postdoctoral Fellowship.

**Authors contributions**

P.P. designed and built the microscope and collected data, J.S. and L.S. designed and implemented the software, S.P. helped with processing the image data, M.W. collected data and designed SolidWorks concepts, K.E. & J.S. co-supervised the software part of the project, J.H. designed the setup, P.T. conceived and supervised the project, collected data, performed data analysis and wrote the paper. All authors contributed to the wiki.

**Figures and Figure legends**

# Figure 1

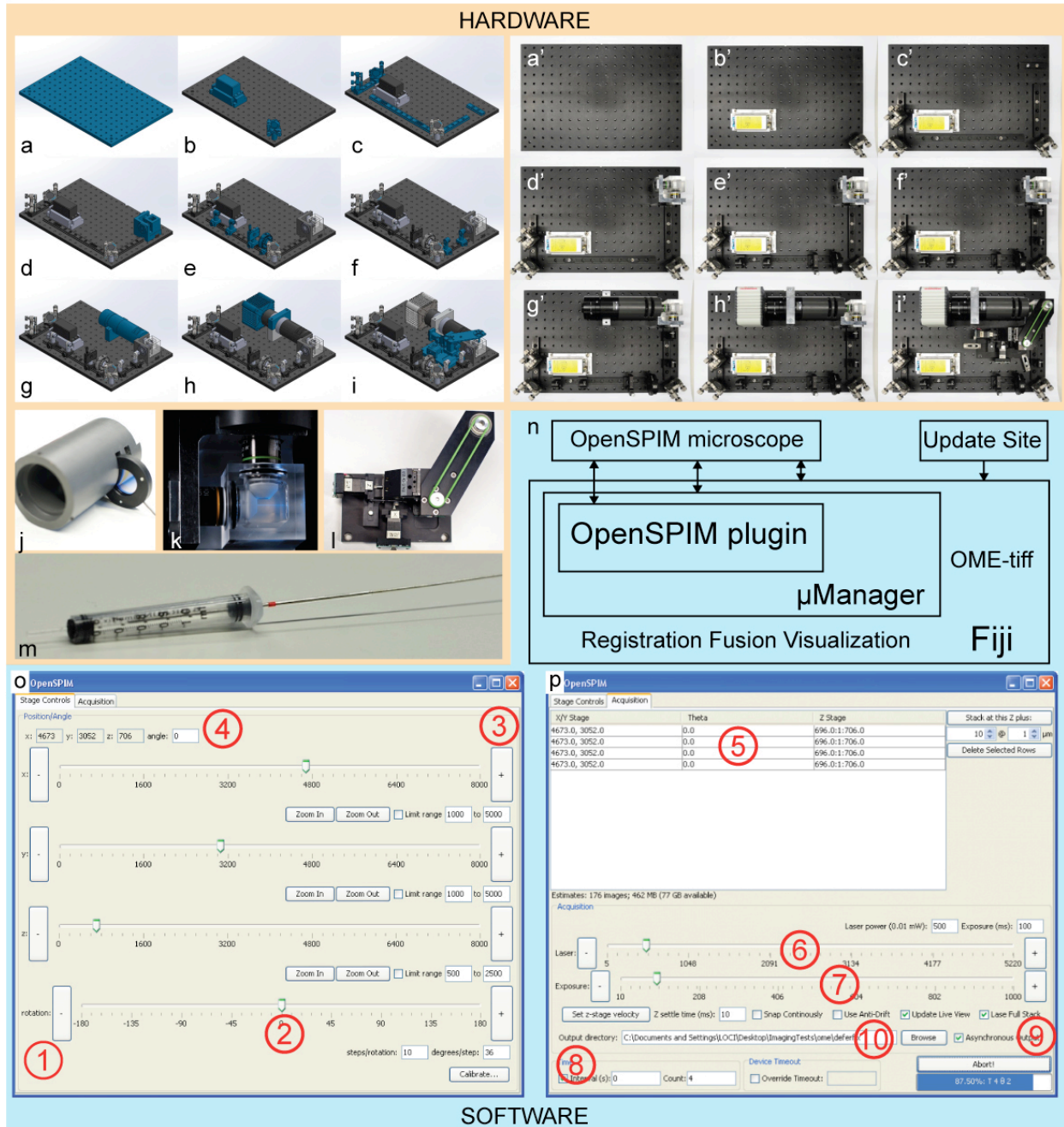

**Figure 1** *OpenSPIM hardware and software* (**a-i**) SolidWorks rendering of the steps necessary to assemble the OpenSPIM, (**a'-i'**) equivalent stages of assembly in real photographs. (**a,a'**) Optical breadboard of the OpenSPIM setup, (**b,b'**) laser on a custom made heat sink and corner mirror, (**c,c'**) dovetail optical rails and mirrors that invert the laser beam, (**d,d'**) OpenSPIM sample chamber, (**e,e'**) light sheet formation

optics including two beam expander lenses, vertical slit and cylindrical lens, (**f,f'**) telescope optics to image the back focal plane of the illumination objective onto an adjustable mirror, (**g,g'**) infinity space tube, tube lens and camera mount, (**h,h'**) CCD camera, (**i,i'**) 4D USB motor system. Newly added components in (**a-i**) are colored blue. (**j**) Infinity space tube with slits for emission filters mounted in custom made holders. (**k**) Sample chamber consisting of Olympus 10x/0.3 illumination and 20x/0.5 detection objectives arranged perpendicularly in a custom made metal (outer) and acrylic (inner) sample chamber. The chamber is filled with water and the light sheet is on. (**i**) 4D USB sample positioning system with sample holder arm and pulley system translating the movement of the fourth motor into rotation. (**m**) Modified plastic syringe serving as a simple holder for glass capillary with specimen in agarose. (**n**) Schematic representation of the OpenSPIM steering software architecture. Data from the OpenSPIM microscope are collected by the OpenSPIM plugin that extends microManager running in Fiji where the data are available for reconstruction and analysis. The design enables active feedback of on-the-fly image processing results on image acquisition on multiple levels (bidirectional arrows). The software is loaded from a dedicated update site and the data are stored using the OME-tiff standard. (**o**) Screenshot of the 4D (xyz and rotation (1)) stage control of the OpenSPIM plugin; sample can be moved by dragging the slider (2), clicking the +/- buttons (3) or typing in the position (4). (**p**) Screenshot of multi-view time-lapse setup window of the OpenSPIM plugin; the microManager position list includes rotation settings (5), laser power (6) and exposure time (7), delay and number of time points (8) can be set; laser can be on for the duration of stack or synchronized with the camera (9), anti-drift option (10) uses image processing and active feedback to motors to keep the sample in the field of view.

# Figure 2

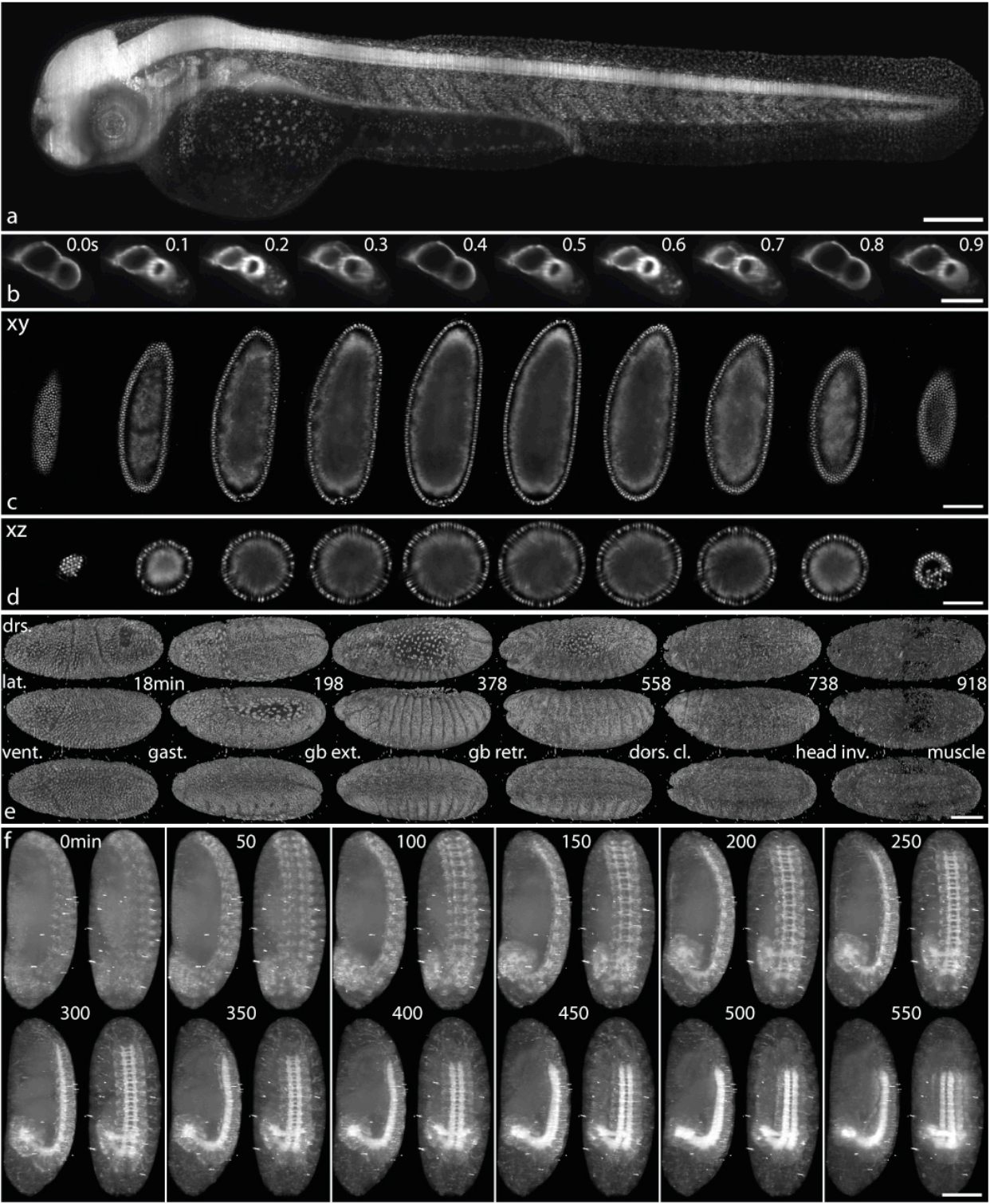

**Figure 2** *OpenSPIM data* (**a**) Two days old zebrafish larva expressing H2A-GFP under the control of beta-actin promoter, Tg(ßactin:H2A-EGFP), was imaged as a set of 6

overlapping fields-of-view montaged using Fiji's stitching plugin with maximum intensity fusion method. (**b**) Beating heart of two days old zebrafish larva expressing the cardiac myosin light chain eGFP fusion Tg(cmlc2:EGFP), serving as myocardium specific marker, was imaged with single plane illumination at 10 frames a second. Two full heart beats are captured at this rate. (**c**) Blastoderm stage *Drosophila* embryo expressing His-YFP in all cells imaged from 6 views at 6 μm steps of the light sheet. The data were reconstructed using Fiji's bead based registration algorithm and fused using multi-view deconvolution. Every 28th plane from the reconstructed volume in xy direction is shown while (**d**) shows every 77th plane from the same volume in the orthogonal direction along the rotation axis. (**e**) 3D rendering of *Drosophila* embryos, expressing His-YFP in all cells, imaged from 5 angles every 6 minutes from gastrulation until hatching. Dorsal (dors.), lateral (lat.) and ventral (vent.) views are shown for every 30th timepoint highlighting the major morphogenetic transition in embryogenesis (gastrulation (gast.), germ band extension (gb. ext.), germ band retraction (gb. retr.), dorsal closure (dors. cl.), head involution (head inv.) and the onset of muscle contraction (muscle)). (**f**) *Drosophila* embryos expressing Csp-sGFP protein fusion under native promoter control imaged from 5 views every 10 minutes. Maximum intensity projection along the lateral and dorsal-ventral axis are shown for every 5th timepoint from germband retraction stage until late embryogenesis highlighting the dynamic morphogenetic movement of the nervous system. Scale bar in all panels is 100 μm.

# Figure 3

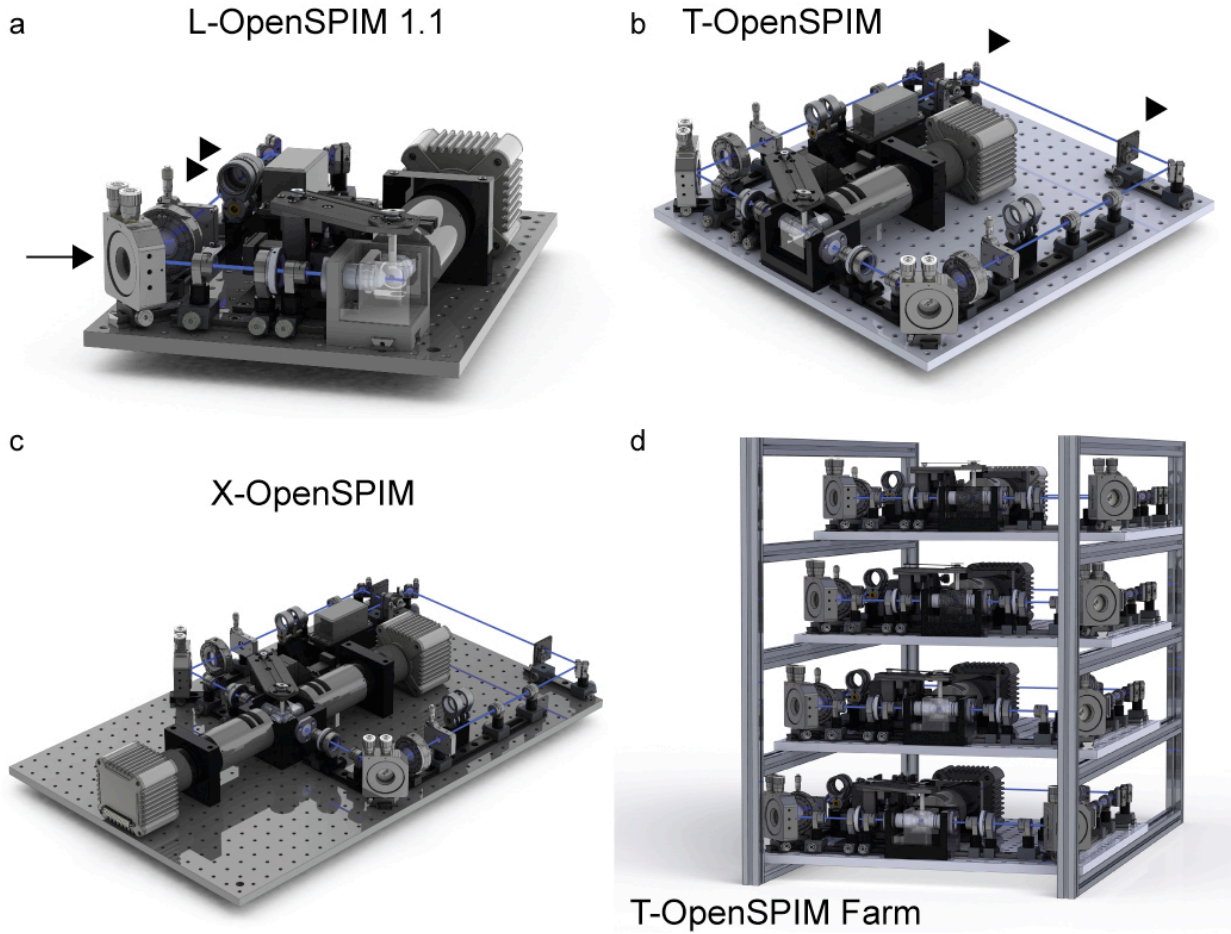

**Figure 3** *OpenSPIM concept configurations* (**a**) OpenSPIM 1.1 - a small, incremental evolution of the current OpenSPIM 1.0 setup adding a gimbal corner mirror mount (arrow) for better control of light sheet focussing and removable neutral density and cleanup filters (arrowheads). One-sided illumination and one-sided detection form an L-OpenSPIM (**b**) Two-sided illumination and one-sided detection T-OpenSPIM configuration with single laser line and shutters for both illumination arms (arrowheads). (**c**) X-OpenSPIM with two-sided illumination and two-sided detection. (**d**) T-OpenSPIM farm of four parallel setups mounted in a rack with a single laser serving the left and right illumination arm of each unit. The racks can be moved individually for convenient sample chamber access.

**Supplementary Video 1** *Stitched zebrafish larva.* Sweep through the 3D volume of two-day-old living zebrafish larva mounted in 1% agarose expressing H2A-GFP under the control of ß-actin promoter (Tg(ßactin:H2A-EGFP) in all cells imaged as a set of 6 overlapping fields-of-view of 80 slices 3 µm apart. The stacks were montaged using Fiji's stitching plugin with maximum intensity fusion method.
http://openspim.org/File:Supplementary_Video_1.ogv

**Supplementary Video 2** *Beating zebrafish heart.* Beating heart of a two-day-old zebrafish larva expressing the cardiac myosin light chain eGFP fusion Tg(cmlc2:EGFP), serving as myocardium specific marker, was imaged with single plane illumination at 10 frames a second (maximum of the Hamamatsu Orca ER camera). Two full heart beats per second are captured at this rate. On the left side the fluorescence of the heart reporter is captured together with the brightfield image revealing the anatomical context.
http://openspim.org/File:Supplementary_Video_2.ogv

**Supplementary Video 3** *Deconvolved blastoderm stage* Drosophila *embryo.* Sweep through the reconstructed volume of a blastoderm stage *Drosophila* embryo expressing His-YFP in all cells imaged from 6 views at 6 µm steps of the light sheet. The data were reconstructed using Fiji's bead based registration algorithm and fused using multi-view deconvolution (Preibisch et al. manuscript in preparation) for 12 iterations. The left embryo shows the data from the orientation of the first acquired view, the middle embryo shows the same volume after 90° rotation along the anterior posterior axis (note that data were not acquired along this exact orientation and yet the quality is very similar). The embryo on the right side shows axial resolution of the reconstructed volume approximately orthogonal to the rotation axis.
http://openspim.org/File:Supplementary_Video_3.ogv

**Supplementary Video 4** *Content-based fused blastoderm stage* Drosophila *embryo.* Sweep through the reconstructed volume of a blastoderm stage *Drosophila* embryo expressing His-YFP in all cells imaged from 6 views at 6 µm steps of the light sheet. The data were reconstructed using Fiji's bead based registration algorithm and fused

using the content-based fusion plugin with blending between views. The left embryo shows the data from the orientation of the first acquired view, the middle embryo shows the same volume after 90 degrees rotation along the anterior posterior axis (note that data were not acquired along this exact orientation and yet the quality is very similar). The embryo on the right side shows axial resolution of the reconstructed volume roughly orthogonal to the rotation axis. Note the significant increase of the background blur compared to **Supplementary Video 3** showing fusion by multi-view deconvolution.
http://openspim.org/File:Supplementary_Video_4.ogv

**Supplementary Video 5** *3d rendering of* Drosophila *embyrogenesis captured by OpenSPIM.* A *Drosophila* embryo expressing His-YFP in all cells was imaged from 5 views every 6 minutes (acquisition of the 5 views took 3 minutes and 30 seconds at 1 mW laser power, 100 ms exposure time and 50 slices 6 µm apart per view) from gastrulation until hatching. The multi-view data were reconstructed using bead based registration and fused with multi-view deconvolution for 15 iterations. A macro script exploiting Fiji's 3D Viewer was used to render the reconstructed volume from each time-point from dorsal (top), lateral (middle) and ventral (bottom) viewpoints. Note the residual beads around the sample which come from enhancement of the weak red fluorescent bead signal by the deconvolution procedure. The first 198 time-points of 235 time-point time-lapse are visualized.
http://openspim.org/File:Supplementary_Video_5.ogv

**Supplementary Video 6** *Multi-view time-lapse of the expression pattern of* Csp. A *Drosophila* embryo expressing Csp-sGFP protein fusion under native promoter control was imaged from 5 views every 10 minutes (acquisition of 5 views took 4 minutes 30 seconds at 1 mW laser power, 500 ms exposure time and 50 slices 6 µm apart per view). On the left side maximum intensity projection along the lateral and dorsal-ventral axis are animated from germband extension stage until late embryogenesis highlighting the dynamic morphogenetic movement of the nervous system. Csp is expressed in all epidermal cells localized to the membrane and this signal dominates during earlier time-points of the movie. Over time the neuronal signal increases. The blur towards the end

of the series is caused by the movement of the living embryo. The maximum intensity projection on the top right side, roughly alongside the rotation axis shows that despite the comparatively low resolution the tissue level expression pattern can be discerned. Bottom right shows 3d rendering of Csp signal over time using a fixed threshold that isolates the stronger nervous system signal revealing striking relocation of the brain hemispheres during head involution.

http://openspim.org/File:Supplementary_Video_6.ogv